\documentclass[useAMS,usenatbib]{mn2e}
\usepackage{graphicx}

\def\lea{\mathrel{<\kern-1.0em\lower0.9ex\hbox{$\sim$}}}
\def\gea{\mathrel{>\kern-1.0em\lower0.9ex\hbox{$\sim$}}}
\def\leq{\mathrel{<\kern-1.0em\lower0.9ex\hbox{$-$}}}
\def\geq{\mathrel{>\kern-1.0em\lower0.9ex\hbox{$-$}}}

\title[Optical/IR Photometry of Open Clusters]{WIYN Open Cluster Study XVI: 
Optical/Infrared Photometry and Comparisons With Theoretical Isochrones}

\author[A. J. Grocholski and A. Sarajedini]
{Aaron J. Grocholski\thanks{aaron@astro.ufl.edu} and Ata 
Sarajedini\thanks{ata@astro.ufl.edu}\\
Department of Astronomy, University of Florida, P.O. Box 112055, Gainesville, 
FL 32611, USA}

\begin{document}
\label{firstpage}
\maketitle

\begin{abstract}
We present combined optical/near-IR photometry (BVIK) for six open clusters -
M35, M37, NGC 1817, NGC 2477, NGC 2420, and M67. The open clusters span an
age range from 150 Myr to 4 Gyr and have metal abundances from [Fe/H] = --0.3
to +0.09 dex. We have utilized these data to test the robustness of
theoretical main sequences constructed by several groups as denoted by the
following designations - Padova, Baraffe, Y$^2$, Geneva, and Siess. The
comparisons of the models with the observations have been performed in the
[$M_V, (B-V)_0$], [$M_V, (V-I)_0$], and [$M_V, (V-K)_0$] colour-magnitude
diagrams as well as the distance-independent $[(V-K)_0, (B-V)_0]$ and
$[(V-K)_0, (V-I)_0]$ two-colour diagrams. We conclude that none of the
theoretical models reproduce the observational data in a consistent manner
over the magnitude and colour range of the unevolved main sequence. In
particular, there are significant zeropoint and shape differences between the
models and the observations. We speculate that the crux of the problem lies
in the precise mismatch between theoretical and observational
colour-temperature relations.  These results underscore the importance of
pursuing the study of stellar structure and stellar modelling with even
greater intensity.
\end{abstract}

\begin{keywords}
Hertzsprung-Russell (HR) Diagram - open clusters and associations: general - stars: evolution - stars: fundamental parameters - stars: late-type
\end{keywords}

\section{Introduction}

With the possible exception of the white dwarf cooling sequence, the main 
sequence is perhaps the best understood phase of stellar evolution.
Conversion of hydrogen to helium occurs in the stellar core representing a
stable and long lasting phase of evolution. As we proceed along the main
sequence from high mass stars to low mass, the production of energy in the
core transitions from the CNO bi-cycle to the proton-proton chain, while,
at the same time, the primary energy transfer mechanism in the core goes from
convection to radiative diffusion. In addition, the outer convection zone
deepens progressively until the star becomes fully convective below
$\sim$0.35$M_\odot$ (Baraffe et al. 1998). All of these features are modelled
with varying degrees of success in modern stellar structure and evolution
codes.

The {\it degree} of success we attribute to stellar models is primarily based
on how well they match the observations. Comparisons such as this have been
performed in a number of ways. One method that has been used in the past
involves comparing isochrones to the locations of nearby field
stars in the HR Diagram (e.g. Baraffe et al. 1998; Bergbusch \& VandenBerg 2001; 
Yi et al. 2001).
Aside from the uncertainties inherent in their trigonometric parallaxes, the
primary drawback of using these stars to test the isochrones is the need to
correct their colours for their differing metallicities.  

Another method that has been used to test the theoretical isochrones is
exemplified by the work of Andersen et al. (1991), Nordstr\"{o}m \& Johansen
(1994), and most recently by Lastennet \& Valls-Gabaud (2002, and references
therein). These authors have determined the masses, effective temperatures,
and luminosities of individual stars in nearby binary systems and compared
their HR diagram locations to the predictions of theoretical models. While
the method has a number of intrinsic advantages, its primary shortcoming is
the obvious requirement that stars must be in binary systems and in relative
proximity to the Sun to be studied in detail.

A third method used to test theoretical stellar models, and the one we are
utilizing in the present work, involves comparing the theoretical isochrones with
{\it cluster} colour-magnitude diagrams (CMDs) and two-colour diagrams (TCDs)
in various passbands (e.g. Castellani et al. 2002; von Hippel et al. 2002; 
von Hippel, Sarajedini, \& Ruiz 2003). Since all stars
are of the same age and metal abundance, this approach is less uncertain than
comparing with individual field stars; however, the reddening is still required to
perform the comparisons in the TCD (though the similarity in the slopes of
the reddening vector and the model loci mitigates this somewhat), 
and both the distance and reddening are needed for the CMD comparisons. 
With the veritable explosion of recent high quality CCD photometry for open 
clusters, this technique is becoming increasingly more powerful as we will
demonstrate in the present work.

This paper is organized as follows.
Section 2 presents a discussion of the optical and IR photometry.  We briefly compare
and contrast the various stellar evolution models in Section 3.  Section 4
represents the primary scientific results of the present investigation;
herein, we discuss the successes and failures of the models in terms of how
well they match the two-colour diagrams and colour-magnitude diagrams of 
open clusters. Finally, we summarize our results in Section 5.

\section{Observational Data}

The optical photometry for the six open clusters considered herein is taken
from the following sources: M35, NGC 2420, NGC 1817 (WIYN Open Cluster Study,
Deliyannis et al. in preparation), M67 (Montgomery, Marschall, \& Janes 1993), M37
(Kalirai et al. 2001), and NGC 2477 (Kassis, et al. 1997).  To
complement the optical data, we utilize $JHK_S$ infrared photometry available
from the Second Incremental Data Release of the 2MASS Point Source
Catalog\footnote{http://irsa.ipac.caltech.edu/}.  For all clusters in our
sample we use the same criteria for the 2MASS data retrieval.  First, source
brightness is limited to 6th magnitude or fainter due to the possible
saturation of bright sources (Carpenter 2000). Second, we have initially
chosen all cluster samples to have a radius of 30 arcmin. The radius is then
reduced to as small as 10 arcmin so as to isolate cluster members and minimize
field star contamination.  Lastly, the 2MASS catalogue provides a read flag
(rd\_flg) which indicates how photometry of each star was measured.  We have
excluded from our sample any sources marked with a read flag of zero in any
one band since this imples that the star was not detected in that band and
the magnitude reported is therefore only an upper limit.  We note that a
majority of source magnitudes come from point spread-function fitting
(rd\_flg = 2), however, we have chosen to include stars with aperture
photometry (rd\_flg = 1) so as to allow the inclusion of brighter cluster
stars.

The 2MASS program uses a K-short ($K_S$) filter for their photometry, which we
have chosen to convert to the Bessell \& Brett (1998, hereafter BB) K-band using
the transformation equations derived by Carpenter (2001) adapted as
follows:
\begin{equation}
(J-K_{BB}) = [(J-K_S) - (-0.011 \pm 0.005)] / (0.972 \pm 0.006)
\end{equation}
and
\begin{equation}
K_{BB} = [K_S - (-0.044 \pm 0.003)] - (0.000 \pm 0.005)(J-K_{BB}).
\end{equation}
The BB photometric system has been adopted by a majority of the stellar 
evolutionary models analysed in the present paper.

We combine optical and infrared photometry for stars in each cluster that are
common to both data sets. All photometry files have RA and Dec available for
each star.  To combine the data, we calculate the spatial offset between each
star in the IR dataset with every star listed in the corresponding optical
dataset.  When the spatial offset is minimized to within a given threshold,
the two stars are considered a match.  The threshold distance is chosen based
on the combined RA and Dec errors and turns out to be 2.8 arcsec for all clusters.

Twarog et al. (1997) have compiled a database of information for 76 open
clusters, for which they provide internally consistent values of reddening,
distance modulus, and metallicity.  All of the clusters discussed in this
study are included in the Twarog et al. (1997) catalogue, so we choose to adopt
their values for the reddening and metallicity. Cluster
ages are taken from Grocholski \& Sarajedini (2002). In
the case of M35, its age is calculated using the method described in
Grocholski \& Sarajedini (2002), which involves adopting ages from the
WEBDA\footnote{ http://obswww.unige.ch/webda/webda.html} 
database and offsetting them to the age scale of Sarajedini (1999).   
We note that while we have tried to make the cluster ages as internally 
consistent as possible, small variations in age have no effect on the location of 
theoretical isochrones along the unevolved MS.  
Table 1 lists these cluster parameters along
with the filter passbands of the available photometry.

Since we are primarily interested in comparing theoretical isochrones to cluster 
photometry in a variety of passbands,
we have chosen to determine the distance moduli in the following manner. 
Given the
adopted reddenings in Table 1, we shift each set of isochrones (described in Sec. 3)
vertically in the $(V-I)_0$ CMD until the model main sequence at $M_V$=6.0 fits
that of the clusters.  Since M37 does not have I-band data, we utilize $(V-K)_0$ 
for the MS fitting.  We have avoided using $(B-V)_0$ for this purpose because it is
more sensitive than $(V-I)_0$
to metallicity variations at temperatures above $\sim$ 4000K ($M_V \sim$ 8.5) which could 
result in erroneous distance determinations in cases where the closest available 
isochrone metallicity differs somewhat from the actual cluster abundance (Baraffe
2003, private communication).  The results, along with the distances given by
Twarog et al. (1997), are given in Table 2. In this way, we can be sure that the models
fit the main sequences in one colour at a given point allowing us to investigate how well 
they fit in other colours at other points along the MS.

To correct for interstellar extinction, we adopt the reddening law derived by
Cardelli, Clayton, \& Mathis (1989), which, using their value of $R_V$ = 3.1
and taking into account the central wavelength of each filter, we find the
following relations for the near-IR passbands: $A_J = 0.282 A_V$, $A_H =
0.180 A_V$, $A_K = 0.116 A_V$.  For observations of M 37, $A_B = 1.41 A_V$
and for all other optical photometry, $A_U = 1.57 A_V$, $A_B = 1.35 A_V$,
$A_R = 0.839 A_V$, and $A_I = 0.560 A_V$.  

In Figs. 1 and 2 we present [$M_V$, $(B-V)_0$] and [$M_V$,
$(V-I)_0$] CMDs, respectively, for all clusters using the full optical data
and in Fig. 3 we show [$M_V$, $(V-K)_0$] CMDs using the combined optical
and infrared datasets.  The clusters are plotted in order of increasing age from
upper left to lower right. We remind the reader that M 37 does not have I band observations
available and is not included in Fig. 2.

\section{Theoretical Models}

We now introduce and briefly compare the five sets of theoretical isochrones that
are used in the present analysis -- Padova (Girardi et al.  
2002)\footnote{http://pleiadi.pd.astro.it/~lgirardi/isoc\_photsys.html}, Baraffe
(Baraffe et al.  1998)\footnote{see reference for anonymous ftp instructions},
Y$^2$ (Yi et al. 2001)\footnote{http://www.astro.yale.edu/demarque/yyiso.html},
Geneva (Lejeune \& Schaerer
2001)\footnote{http://webast.ast.obs-mip.fr/stellar/}, and Siess (Siess,
Dufour, \& Forestini
2000)\footnote{http://www-laog.obs.ujf-grenoble.fr/activites/starevol/evol.html}.  
While we do not intend this to be a comprehensive review of the stellar models,
we point out some of the similarities and differences between the input parameters
of the five groups as  listed in Table 3.

With respect to the opacities, all of the groups use versions of
the OPAL opacities (e.g. Iglesias \& Rogers 1996) for high temperatures
supplemented by the Alexander \& Ferguson (1994, AF94) tables for low
temperatures.  The only caveat is that the Geneva models also consider the
Kurucz (1991) low temperature opacities.  We see that all five groups utilize
different sources for their equations of state.  Three of the groups use
similar values for core overshoot in all of their models, while Siess has
only included overshoot for their solar metallicity models.  Baraffe has
chosen to not include core overshoot because the focus of
their work is low mass stars, which, for masses below $\sim
0.35M_{\odot}$, are fully convective so an overshoot parameter is not
necessary.  Also, for low mass stars above this limit, a convective core is
only present for a small fraction of time during their evolution, so
inclusion of overshoot will not have a significant effect on these objects (Chabrier
\& Baraffe 1997).  Mixing length parameters are similar for all groups,
varying by less than 20\%.  It is important to keep in mind that the Baraffe 
models are available
with three mixing length values, $\alpha = $ 1.0, 1.5, 1.9.  Setting the
mixing length equal to 1.9 is required to match observations of the Sun
(Baraffe et al. 1998), however these models are not extended below
$0.6M_{\odot}$ ($M_V \sim 8.8$).  Baraffe et al. (1998) find that, for $M
\leq 0.6M_{\odot}$, their isochrones with the three different mixing lengths
converge, implying that the mixing length has little effect on the models
below this mass.  As such, we combine the $\alpha = 1.9$ models for masses
above $\sim 0.6M_{\odot}$ with $\alpha = 1.0$ models covering the very low
mass stars.

Lastly, we compare each group's method of deriving synthetic photometry for
their models which is then used to convert effective temperatures to colours.  
The calculations by both the Geneva and Y$^2$ groups come from
versions of the $BaSeL$ stellar spectral library (see eg. Lejeune and
Schaerer 2001).  This library is based on model atmosphere spectra by Kurucz
(1995), Fluks et al. (1994), and Bessell et al. (1989, 1991).  Padova
calibrations are also based largely on the spectral library of Kurucz (1992)  
with extension to cool stars through the use of Fluks et al. (1994) empirical
M giant spectra and the DUSTY99 (Chabrier et al. 2000) atmosphere models of
cool dwarf (type M and later) synthetic spectra (see $\S$3 of Girardi et al.
2002 for a complete discussion).  The Siess model uses $T_{eff} - BC$
relations from Schmidt-Kaler (1982) and Bessell (1991) and $T_{eff} - colour$
relationships from Fitzgerald (1970) and Arribas \& Martinez-Rogers (1988,
1989).  Lastly, the Baraffe isochrones use stellar spectra generated with the
NextGen model atmosphere code (eg. Hauschildt, Allard, \& Baron 1999)  
which is a predecessor of the aforementioned DUSTY99 models.

In Fig. 4 we plot log$(L/L_{Sun})$ vs. log$(T_{eff})$ for each of the
models used in this study (note that we have included the complete $\alpha
= $ 1.0 as well as the $\alpha = $ 1.9 Baraffe models).  All isochrones are
for solar metallicity and an age of 5 Gyr; the position of the Sun is
indicated by the filled circle.  The inset box shows a close-up of the 
region around the Sun. Comparing the models in the theoretical
plane allows us to eliminate the uncertainty inherent in the transformation to the observational plane.  In this figure we see
that four of the plotted isochrones, Padova (solid line), Baraffe, $\alpha =
1.9$ (dotted line), Y$^2$ (dash-dot line), and Geneva (dash-dot-dot line)
are all in good agreement for the upper MS and the position of the
Sun, but show considerable deviation beginning around $0.1L_{\odot}$ 
($M_V$$\sim$7.5, $0.65M_{\odot}$, $T_{eff} \sim 4100$K, and spectral 
type $\sim$ K7).  Neither Siess nor
Baraffe ($\alpha = 1.0$) match the Sun and, for a given luminosity, predict
cooler effective temperatures than the other isochrones on the upper MS.

Figure 5a displays the relationship between 
model mass and absolute magnitude in the V-band. In particular, the lower panel 
of Fig. 5a shows the difference in mass between each isochrone and the Padova set, 
which has been chosen as an arbitrary reference. We see that the mass differences
are minimized at $M_V$$\sim$6.0, the magnitude at which we have determined the 
distance modulus of each cluster (Sec. 2). At fainter magnitudes
the differences between the various models steadily increase but are generally
less than 0.1$M_{\odot}$. Fainter than $M_V$$\sim$9, the masses of the $Y^2$ 
models begin to deviate from the others. As we discuss how well the models fit the
observations in the CMD, Fig. 5a will prove useful in estimating the mass at each
value of the absolute magnitude.  

Figure 5b compares the difference in colour between each model and the Solar abundance Padova isochrone as a function of absolute V-band magnitude for $(B-V)_0$ (top
panel), $(V-I)_0$ (middle), and $(V-K)_0$ (bottom).  As with Fig. 5a,
the colour differences are generally smaller at $M_V$$\sim$6.0
and increase at fainter magnitudes. Furthermore, one would expect that
if the models all used the same method for converting effective
temperatures to colours, then the appearance of Fig. 5b would be
solely a reflection of the
differences in the physics of the models (see Fig. 4).  We see,
however, that this is not the case; there are no clear systematic 
trends in these differences from model to model (with the possible 
exception of the
Geneva isochrones).  These virtually random differences indicate that
the disagreement between models is a result of both the chosen physics and 
the colour transformations.

\section{Results}

\subsection{Color-Magnitude Diagrams}

As mentioned in $\S1$, we are interested in comparing the observational CMDs
with theoretical models for a variety of colours.  These comparisons are illustrated
in Figs. 6 through 22\footnote{Only a sample of these CMDs (Figs. 6, 12, and 17) are included in the hard copy of this paper. The rest are only available in the electronic copy.} for the 6 clusters in our sample. In each case, we show
the [$M_V, (B-V)_0$], [$M_V, (V-I)_0$], and [$M_V, (V-K)_0$] diagrams along
with the isochrone that is closest to the clusters' metallicity as listed in Table 1.
Listed in each plot are the cluster name and
model name, along with two numbers that indicate which isochrone is used.  
The first number gives the log age of the model (e.g.  0945 $\Rightarrow$
log$(Age)  = 9.45$) and the second indicates the metal abundance (e.g. z019
$\Rightarrow$ Z = 0.019).
In the case of the Padova isochrones, Leo Girardi kindly provided interpolated
models that match the cluster metal abundances. In addition, because the M37 
data of Kalirai et al. (2001) does not possess I-band photometry, we only show 
data in the $(B-V)_0$ and $(V-K)_0$ colours. To simplify the discussion below, we will
refer to mismatches between the models and the observations as occuring in the
magnitude direction with the understanding that there may be a component
in the colour direction as well.

Looking at the CMD and isochrone comparisons in Figs. 6 through 22, a number
of trends stand out. First, we confirm that all of the models fit the cluster MS
at $M_V$=6.0 in the $(V-I)_0$ plane. This is to be expected given the
manner in which we determined the cluster distance moduli. Next,
we examine the isochrone comparisons in the $(B-V)_0$ plane shown in Figs. 6
through 11. In the vast majority of cases, the model main sequences match
the observations at $M_V$=6.0 adequately. There are two notable exceptions to this. 
First, the M37 photometry is significantly fainter than all of the isochrones 
considered herein by between $\sim$0.5 and $\sim$0.9 mag in colour. 
As such, we will exclude it from further discussion within the context of the other
$(B-V)_0$ CMDs. Second, the Baraffe models are consistently brighter 
than the observations at $M_V$=6.0, except in the case of M35. Moving on to 
the other regions of the CMD in the $(B-V)_0$ plane, all of the models reproduce the
unevolved MS in the range  2.0$\lea$$M_V$$\lea$7.0 reasonably well. 
The primary exception to this is the fit of the Geneva models to the M35 
photometry shown in the middle left panel of Fig. 6. In this case, the
MS of the isochrone is clearly offset from the observed one for
$M_V$$\lea$5.0. Another exception, though not as obvious, is apparent in the
comparison between the Y$^2$ models and the M35 data in the upper right
panel of Fig. 6. Again, the isochrone is brighter than the data in the range
2.0$\lea$$M_V$$\lea$5.0. Thirdly, there is the comparison between NGC 2420 and
the Siess models in Fig. 10 where the theoretical MS lies above the data. 
It is unclear what these mismatches are due to.
However, in the case of the Geneva models, it may be a result of the fact that
the cluster (M35) metallicity is lower than those used in the isochrones. For the clusters
whose photometry extends reliably below $M_V$$\sim$8.0 (NGC 2477 and M67, 
again excluding M37), we find that the Padova and Baraffe models diverge the most from the observed
main sequences while the Y$^2$ and Siess models deviate the least. Note that the 
Geneva model MS does not extend faint enough for this comparison to be made.

Moving on to a consideration of the [$M_V, (V-I)_0$] diagrams in Figs. 12
through 16, we see a similar behaviour in the range 2.0$\lea$$M_V$$\lea$7.0 as
noted in the case of the $(B-V)_0$ CMDs. In short, all of the models perform
reasonably well in matching the observed main sequences. The difficulties arise
in the magnitude ranges fainter than $M_V$$\sim$8.0; the Padova models
diverge the most from the cluster main sequences, while the Baraffe
theoretical main sequences reproduce the observed ones more consistently 
than any of other sets.

Lastly, we examine the  [$M_V, (V-K)_0$] diagrams which were constructed
by matching optical photometry with 2MASS near-IR photometry. 
We note that the level of agreement between the isochrones and the data at 
$M_V$=6.0
is not consistent from cluster to cluster and model to model. With the exception
of M37 where the distance fits have been performed in the [$M_V, (V-K)_0$] plane,
in most cases, the model
main sequences are fainter than the observed ones. 
However, the Padova MS matches that of M35 at 
$M_V$=6.0; similarly, the Geneva models match the MS of NGC 2477 at
$M_V$=6.0. In terms of the level of overall agreement
along the MS, we can say that, like the $(B-V)_0$ and $(V-I)_0$ CMDs,
the $(V-K)_0$ models perform better in the range 2.0$\lea$$M_V$$\lea$7.0
as compared to fainter magnitudes. However, with the exception of M35, the
models are generally fainter than the observational main sequences.

To conclude this section, we note that {\it none}
of the theoretical models fit the unevolved MS (i.e. $M_V$$\gea$4) 
adequately in all three colours. Either the shape of the isochrone does not 
match the observations or the location does not match or both. The small difference
between the metallicity of the isochrones and the observations is unlikely to be 
the cause of this mismatch.  

\subsection{Two Color Diagrams}

One of the ways in which we can eliminate the use of the distance modulus 
in the present analysis is to compare the optical/IR photometry of open clusters
with theoretical isochrones in the two-colour diagram (TCD). In this plane, only the
reddenings of the clusters are required allowing us to further investigate the
robustness of the models. Figures 23 through 28 show the $[(V-K)_0, (B-V)_0]$
TCDs while Figs. 29 through 33\footnote{Again, only a sample of these figures are shown in the hard copy (Figs. 23 and 29) while the remainder are only available electronically.} display the
$[(V-K)_0, (V-I)_0]$ TCDs for main sequence stars in our sample of open clusters. 
The arrow in each figure represents the slope of the reddening vector 
($E(B-V)=E(V-I)=0.3$) which is 
roughly parallel to the two-colour sequence suggesting that inferences made from 
these diagrams are largely insensitive to the adopted reddenings.

Looking first at the $[(V-K)_0, (B-V)_0]$ diagrams, except in a minority of cases,
there is a zeropoint offset between the theoretical two-colour sequence and the
observations; the models are too blue in $(V-K)_0$ at a given
$(B-V)_0$ as compared with the observational data. In addition, the detailed shapes 
of the theoretical 
sequences can be dramatically different than the data especially in the case of the
Baraffe models for $(V-K)_0$$\gea$3.0. The M37 and M67 data, which extend to the 
reddest colours, nicely 
underscore these shape differences. Largely similar conclusions (i.e. models are offset to 
bluer colours and don't reproduce the shape of the observed TCDs) can be reached by 
examining the $[(V-K)_0, (V-I)_0]$ diagrams.

As a sanity check on the observations, Figs. 34 and 35 present TCDs of each cluster
compared with the sequence for M35 determined by eye using the data in the
upper left panels. A reddening vector equal to $E(B-V)=E(V-I)=0.3$ is also shown in 
each figure. We see that in most cases, the positions and shapes of the two-colour
sequences for each cluster are in good accord with that of M35. In Fig. 34, the
data for M37 are below the M35 line by $\sim$0.1 mag; this may account for the
fact that in the [$M_V, (B-V)_0$] CMDs (Fig. 7), the M37 data is shifted away from the 
isochrones by a significant amount as compared with the data from the other clusters. 
Figure 34 also shows that the M67 two-colour data appears to be slightly offset from the
M35 sequence in the colour range 0.5$\lea$$(B-V)_0$$\lea$1.0. Lastly, in Fig. 35,
we see that the M67 sequence has a slightly different shape as compared with the
M35 $[(V-K)_0, (V-I)_0]$ line. Taken together, the TCDs in Fig. 34 and 35 provide
reassurance that the observational data we have considered herein are internally
consistent.

\section{Conclusion}

We have presented combined optical/near-IR photometry for six open clusters. The
optical data have been obtained as part of the WIYN Open Cluster Study and taken 
from the literature, while the
near-IR observations are from the Second Incremental Data Release of the 
Two Micron All Sky Survey Point Source Catalog. The open clusters span an age
range from 150 Myr to 4 Gyr and have metal abundances from [Fe/H] = --0.27 to
+0.09 dex. We have utilized these data to test the robustness of theoretical main 
sequences constructed by several groups as denoted by the following designations - 
Padova (Girardi et al. 2002), Baraffe (Baraffe et al.  1998), Y$^2$ (Yi et al. 2001), 
Geneva (Lejeune \& Schaerer 2001), and Siess (Siess et al. 2000). The
comparisons of the models with the observations have been performed in
the [$M_V, (B-V)_0$], [$M_V, (V-I)_0$], and [$M_V, (V-K)_0$] colour-magnitude
diagrams as well as the distance-independent $[(V-K)_0, (B-V)_0]$ and
$[(V-K)_0, (V-I)_0]$ two-colour diagrams. The reddenings have been taken from
Twarog et al. (1997) and the distance moduli were determined by adopting these
reddenings and then matching each isochrone to the observed main sequences
at $M_V$=6.0 in the [$M_V, (V-I)_0$] plane ([$M_V, (V-K)_0$] for M37). 

We conclude that none of the theoretical
models reproduce the observational data in a consistent manner over the
magnitude and colour range of the unevolved main sequence. In particular, there are
significant zeropoint and shape differences between the models and the
observations. In the colour-magnitude diagrams, most of the models performed 
reasonably well for stars with
Mass$\gea$0.75$M_{\odot}$, but did not do so for stars with lower masses. 
In the two-colour diagrams, the theoretical sequences are too blue in $(V-K)_0$ 
at a given $(B-V)_0$ and $(V-I)_0$ as compared with the observational data. 
The shapes of the theoretical sequences can also be dramatically different 
as compared with the data. From Figs. 4 and 5 we can say with a fair amount of certainty that the crux of the problem lies in the precise mismatch between theoretical and observational colour-temperature relations.  
Depending on how these relations are derived, one of the problems 
could be missing sources of opacity in the atmospheres of the model stars.
These results underscore the importance of pursuing the study of
stellar structure and stellar modelling with even greater intensity.

\section*{Acknowledgments}
The authors are grateful to Isabelle Baraffe, Pierre Demarque, and Ted von
Hippel
for providing comments on an early version of this manuscript and to Tom
Kehoe for helping with the preparation of the manuscript.  A.G. is
especially
indebted to Isabelle Baraffe for many enlightening discussions.  The
authors are appreciative of Glenn Tiede for his countless contributions.
We thank Con Deliyannis, Kent Montgomery, and Jason Kalirai for
supplying us with electronic copies of their photometry and Leo Girardi
for providing
models for previously unpublished metallicities.  We also acknowledge 
the referee, Bruce Twarog, for comments that greatly improved the
presentation of this paper.  This publication makes use of data products
from the Two Micron All Sky Survey,
which is a joint project of the University of Massachusetts and the
Infrared
Processing and Analysis Center/California Institute of Technology, funded
by the
National Aeronautics and Space Administration and the National Science
Foundation.

\clearpage

\begin{figure}
 \caption{Colour-magnitude diagrams in the [$M_V, (B-V)_0$] plane
  for the six clusters in our dataset. The distance moduli and reddenings are 
  given in Table 1. All plots have the same axis scale and the clusters are 
  presented in order from youngest (top left) to oldest (lower right).}
\end{figure}

\begin{figure}
\caption{Same as Fig. 1 except that the CMDs are in the 
[$M_V, (V-I)_0$] plane.  We note that M 37 
is not included because Kalirai et al. (2001) do not include I-band
photometry in their dataset.}
\end{figure}

\begin{figure}
\caption{Same as Fig. 1 except that the CMDs are in the 
[$M_V, (V-K)_0$] plane. It is evident from the plots that the CMDs are limited
by the depth of the K-band data.}
\end{figure}

\begin{figure}
\caption{We plot log$(L/L_{\odot})$ vs. log$(T_{eff})$ for each 
of the 
models used in this study.  All isochrones are for solar metallicity at an 
age of 5 Gyr.  The position of the Sun is marked by the filled circle. The
inset shows a close-up of the region around the Sun's location.}
\end{figure}

\setcounter{figure}{4}
\begin{figure}
\caption{(a) The upper panel shows the relationship between 
absolute 
magnitude in the V-band and stellar mass in units of solar masses for
each of the models considered herein. In
the lower panel, we have arbitrarily selected the Padova isochrones as the
reference and plot the mass difference between each set of models
and this reference.}
\end{figure}

\setcounter{figure}{4}
\begin{figure}
\caption{(b) In these three panels we plot the difference in colour between each model and the Solar abundance Padova isochrones as a function of absolute V-band magnitude for $(B-V)_0$ (top panel), $(V-I)_0$ (middle), and $(V-K)_0$ (bottom).}
\end{figure}

\begin{figure}
\caption{Comparisons between the [$M_V, (B-V)_0$] CMD of M35 
and
theoretical models from the Padova, Y$^2$, Geneva and Siess groups.  
Note that the cluster distance modulus has been adjusted so that the
models and observations agree at $M_V$=6.0 in the VI CMDs. The Log 
of the plotted
ages are given along with the metal abundance (Z) of the models.}
\end{figure}

\begin{figure}
\caption{Same as Fig. 6 except that the photometry of M37 is shown.}
\end{figure}

\begin{figure}
\caption{Same as Fig. 6 except that the photometry of NGC 1817 is shown.}
\end{figure}

\begin{figure}
\caption{Same as Fig. 6 except that the photometry of NGC 2477 is shown.}
\end{figure}

\begin{figure}
\caption{Same as Fig. 6 except that the photometry of NGC 2420 is shown.}
\end{figure}

\begin{figure}
\caption{Same as Fig. 6 except that the photometry of M67 is shown.}
\end{figure}

\setcounter{figure}{11}
\begin{figure}
\caption{Comparisons between the [$M_V, (V-I)_0$] CMD of M35 and
theoretical models from the Padova, Y$^2$, Geneva, Siess, and Baraffe groups.  
The Log of the plotted ages are given along with the metal abundance (Z) of the 
models.}
\end{figure}

\clearpage
\begin{figure}
\caption{Same as Fig. 12 except that the photometry of NGC 1817 is shown.}
\end{figure}

\begin{figure}
\caption{Same as Fig. 12 except that the photometry of NGC 2477 is shown.}
\end{figure}

\begin{figure}
\caption{Same as Fig. 12 except that the photometry of NGC 2420 is shown.}
\end{figure}  

\begin{figure}
\caption{Same as Fig. 12 except that the photometry of M67 is shown.}
\end{figure}  

\setcounter{figure}{16}
\begin{figure}
\caption{Comparisons between the [$M_V, (V-K)_0$] CMD of M35 and
theoretical models from the Padova, Y$^2$, Geneva, Siess, and Baraffe groups.  
The Log of the plotted ages are given along with the metal abundance (Z) of the 
models.}
\end{figure}  

\begin{figure}
\caption{Same as Fig. 17 except that the photometry of M37 is shown.}
\end{figure}  

\begin{figure}
\caption{Same as Fig. 17 except that the photometry of NGC 1817 is shown.}
\end{figure}  

\begin{figure}
\caption{Same as Fig. 17 except that the photometry of NGC 2477 is shown.}
\end{figure}  

\begin{figure}
\caption{Same as Fig. 17 except that the photometry of NGC 2420 is shown.}
\end{figure}  

\begin{figure}
\caption{Same as Fig. 17 except that the photometry of M67 is shown.}
\end{figure}  

\setcounter{figure}{22}
\begin{figure}
\caption{Two colour diagrams (TCD) using $(B-V)_0$ and $(V-K)_0$ data for
main sequence stars in M35. The solid lines represent the
models indicated in each panel. These plots show that even when 
we remove the uncertainties inherent in the distance moduli, the theoretical 
isochrones still do not reproduce the shape of the observed sequence in
the TCD. We have included a reddening vector with length $(B-V)_0 = 0.3$}
\end{figure}  

\begin{figure}
\caption{Same as Fig. 23 except that we now compare the observational
data for M37 with theoretical models.}
\end{figure}  

\clearpage
\begin{figure}
\caption{Same as Fig. 23 except that we now compare the observational
data for NGC 1817 with theoretical models.}
\end{figure}  

\begin{figure}
\caption{Same as Fig. 23 except that we now compare the observational
data for NGC 2477 with theoretical models.}
\end{figure}  

\begin{figure}
\caption{Same as Fig. 23 except that we now compare the observational
data for NGC 2420 with theoretical models.}
\end{figure}  

\begin{figure}
\caption{Same as Fig. 23 except that we now compare the observational
data for M67 with theoretical models.}
\end{figure}  

\setcounter{figure}{28}
\begin{figure}
\caption{Same as Fig. 23 except that two colour diagrams using $(V-I)_0$ 
and $(V-K)_0$ data are plotted.  The reddening vector shown has length $(V-I)_0 = 0.3$.}
\end{figure}  

\begin{figure}
\caption{Same as Fig. 29 except that we now compare the observational
data for NGC 1817 with theoretical models.}
\end{figure}

\begin{figure}
\caption{Same as Fig. 29 except that we now compare the observational
data for NGC 2477 with theoretical models.}
\end{figure}

\begin{figure}
\caption{Same as Fig. 29 except that we now compare the observational
data for NGC 2420 with theoretical models.}
\end{figure}

\begin{figure}
\caption{Same as Fig. 29 except that we now compare the observational
data for M67 with theoretical models.}
\end{figure}

\setcounter{figure}{33}
\begin{figure}
\caption{Two colour diagrams using $(B-V)_0$ and $(V-K)_0$ data
for all of the clusters in the present study compared with the observed
sequence for M35 (solid line). We have included a reddening vector 
with length $(B-V)_0 = 0.3$}
\end{figure}

\begin{figure}
\caption{Two colour diagrams using $(V-I)_0$ and $(V-K)_0$ data
for all of the clusters in the present study compared with the observed
sequence for M35 (solid line). We have included a reddening vector with 
length $(V-I)_0 = 0.3$}
\end{figure}

\begin{figure}
\caption{The [$M_V, (V-K)_0$]  CMDs for M35 and M67 with the solid
lines representing the fiducial sequences drawn by eye through the highest density
of points. The right panel shows the fiducials for both clusters plotted in
the same plane along with field stars (open circles) from Percival et al. (2003) with close-to-solar abundances and {\it Hipparcos} parallaxes.}
\end{figure}

\clearpage





\begin{table}
\caption{Open Cluster Information \label{tbl-1}}
\begin{tabular}{@{}lcccc} \hline \hline

Name & Available Photometry & Log Age & $E(B-V)$ & $[Fe/H]$ \\
\hline
M 35 (NGC 2168) &$UBVRIJHK_S$ &8.17 &0.19 &$-$0.160\\
M 37 (NGC 2099) &$...BV.....JHK_S$ &8.73 &0.27 &0.089\\
NGC 1817 &$...BVRIJHK_S$ &8.80 &0.26 &$-$0.268\\
NGC 2477 &$UBV...IJHK_S$ &9.04 &0.23 &0.019\\
NGC 2420 &$...BVRIJHK_S$ &9.24 &0.05 &$-$0.266\\
M 67 (NGC 2682) &$UBVRIJHK_S$ &9.60 &0.04 &0.000\\
\hline
\end{tabular}
\end{table}




\begin{table}
\caption{Open Cluster Distance Moduli. \label{tbl-2}}
\begin{tabular}{@{}lcccccc} \hline \hline

Cluster & Padova & Baraffe & Geneva & Y$^2$ & Siess & Twarog et al. \\
\hline
M 35 (NGC 2168) &10.16 &10.41 &9.81 &9.91 &9.96& 10.30\\
M 37 (NGC 2099) &11.55 &11.40 &11.50 &11.35 &11.75& 11.55\\
NGC 1817 &12.10 &12.30 &11.90 &11.85 &12.00 & 12.15\\
NGC 2477 &11.55 &11.60 &11.30 &11.15 &11.45 & 11.55\\
NGC 2420 &12.12 &12.45 &11.95 &11.90 &12.07& 12.10\\
M 67 (NGC 2682) &9.80 &9.80 &9.60 &9.45 &9.65 & 9.80\\
\hline
\end{tabular}
\end{table}

\clearpage

\begin{table}
\caption{Theoretical Model Input Parameters \label{tbl-3}}
\begin{tabular}{@{}lcccccc} \hline \hline
 & Padova & Baraffe & Geneva & Y$^2$ & Siess \\
\hline
Opacity & OPAL (1993)$^{a}$ & OPAL (1996)$^{b}$ & OPAL
(?)$^{c}$ & OPAL (1996)$^{b}$ & OPAL
(1996)$^{b}$ \\
\\
Low temperature & AF94 & AF94 & AF94, & AF94 & AF94 \\
opacity & & & Kurucz (1991) & & \\
\\
Equation of state & $T>10^7$: Kippenhahn$^{d}$ &
SCVH$^{f}$ & Maeder \& & OPAL (1996)$^{b}$ & based on\\
& $T<10^7$: MHD$^{e}$ & & Meynet (1989) & & Pols et al. (1995)\\
\\
Core overshoot & 0.25$H_p$ for & none & $0.2H_p$ for & $0.2H_p$ for & 
$0.2H_p$
for $Z=0.02$ \\
& $M\geq 1.5M_{\odot}$ & &  $M\geq 1.5M_{\odot}$ & $age \leq 2Gyr$ & (all 
others
= 0) \\
\\
Mixing length, $\alpha$ & 1.68 & 1.9 & 1.6 & 1.7431 & 1.605 \\
\\
He abundance & $Y_p = 0.23$ & $Y_{solar} = 0.282$ & $Y_p = 0.24$ & $Y_p =
0.23$ & $Y_p = 0.235$ \\
\\
He enrichment, $\frac{\Delta Y}{\Delta Z}$ & 2.25 & n/a & 2.5 for
$Z>0.02$ & 2.0 & 2.0 \\
& & & 3 for $Z\leq0.02$ & & \\
\\
Synthetic photometry & ATLAS9$^{g}$ & NextGen$^{i}$ &
BaSeL-2.2$^{j}$ & BaSeL-2.2$^{j}$ & Siess et al.
(1997)\\
& DUSTY99$^{h}$ & & & & \\
& Fluks et al. (1994) & & & & \\
\hline
\end{tabular}
$^{a}${Iglesias, \& Rogers (1993)}

$^{b}${Iglesias, \& Rogers (1996)}

$^{c}${Geneva isochrones were published over the course of
several years and as such utilize OPAL opacities from different years.  See
Lejeune \& Schaerer (2001) for more information.}

$^{d}${Kippenhahan, Thomas \& Weigert (1965)}

$^{e}${Mihalas et al. (1990)}

$^{f}${Saumon, Chabrier, \& VanHorn (1995)}

$^{g}${Castelli, Gratton, \& Kurucz (1997)}

$^{h}${Chabrier et al. (2000)}

$^{i}${Hauschildt, Allard, \& Baron (1999)}

$^{j}${Westera, Lejeune, \& Buser (1999)}

\end{table}

\clearpage

\begin{table}
\caption{Cluster Fiducials \label{tbl-4}}
\begin{tabular}{@{}ccccccc} \hline \hline
 & M35 & & & & M67 &  \\
$V$ & & $V-K$ & & $V$ & & $V-K$ \\
\hline
  9.0 & & 0.32 & & 11.0 & & 2.57 \\
  9.5 & & 0.34 & & 11.5 & & 2.44 \\
10.0 & & 0.36 & & 12.0 & & 2.32 \\
10.5 & &	0.40 & &	12.5 & &	2.20 \\
11.0 & &	0.44 & &	12.9 & &	2.09 \\
11.5 & &	0.51 & &	13.0 & &	2.01 \\
12.0 & &	0.66 & &	12.8 & &	1.85 \\
12.5 & &	0.89 & &	12.7 & &	1.74 \\
13.0 & &	1.17 & &	12.65 & &       1.58 \\
13.5 & &	1.45 & &	12.7 & &	1.40 \\
14.0 & &	1.66 & &	12.8 & &	1.33 \\
14.5 & &	1.83 & &	13.1 & &	1.42 \\
15.0 & &	2.01 & &	13.2 & &	1.41 \\
15.5 & &	2.19 & &	13.4 & &	1.38 \\
16.0 & &	2.39 & &	13.5 & &	1.375 \\
16.5 & &	2.64 & &	13.6 & &	1.374 \\
17.0 & &	2.96 & &	13.7 & &	1.385 \\
17.5 & &	3.29 & &	13.8 & &	1.40 \\
18.0 & &	3.59 & &	14.0 & &	1.43 \\
18.5 & &	3.89 & &	14.2 & &	1.48 \\
19.0 & &	4.19 & &	14.4 & &	1.55 \\
19.5 & &	4.44 & &	14.6 & &	1.62 \\
20.0 & &	4.69 & &	14.8 & &	1.69 \\
        & &	        & &	15.0 & &	1.77 \\
        & &	        & &	15.5 & &	1.99 \\
	 & &		 & &	16.0 & &	2.26 \\
	 & &		 & &	16.5 & &	2.54 \\
	 & &		 & &	17.0 & &	2.84 \\
	 & &		 & &	17.5 & &	3.17 \\
	 & &		 & &	18.0 & &	3.50 \\
	 & &		 & &	18.5 & &	3.80 \\
	 & &		 & &	19.0 & &	4.07 \\
\hline
\end{tabular} 
\end{table}

\appendix
\section{Cluster $V-K$ Fiducial Sequences}

Published $(V-K)_0$ fiducial sequences are rare for open clusters; as such, to help fill
an important need, we have derived such sequences for the youngest (M35) and oldest 
(M67) clusters
in our sample. The left and centre panels of Fig. 36 show our adopted fiducial sequences for
M35 and M67, respectively. These have been constructed by-eye through the
highest density of points in the CMD and are given in Table 4. 
The cluster data and the fiducials are plotted 
after accounting for the effects of reddening (Table 1) and distance using the median of the
values in Table 2. The right panel in Fig. 36 shows the two fiducials plotted together to
reinforce the point that the observations of these clusters agree well with each
other even to the limits of the M67 photometry at $M_V$$\sim$9. 

To investigate how well solar-abundance field MS stars compare with these
two clusters, we have combined optical photometry of field stars with
$[Fe/H]$ values within $\pm$0.1 of the Sun and {\it Hipparcos} parallaxes 
from Table 1 of Percival, Salaris, \& Kilkenny (2003) with K-band data for
these same stars from the  Second Incremental Data Release of the 2MASS 
Point Source Catalog. The open circles in the right panel of Fig. 36 represent three
stars that meet these criteria. It is evident that there is good agreement between
the locations of the field stars and the MS locations of M35 and M67.

\end{document}